\documentclass[epj]{svjour}

\usepackage{graphicx}
\usepackage{siunitx}
\usepackage{amsmath}
\usepackage{amssymb}
\usepackage[version=4]{mhchem}
\usepackage{hyperref}
\hypersetup{
	colorlinks=true,
    linkcolor=blue,
    filecolor=blue,
	urlcolor=blue,
    citecolor=blue,
}
\usepackage{soul}

\newcommand{\orcid}[1]{\href{https://orcid.org/#1}{\includegraphics[width=8pt]{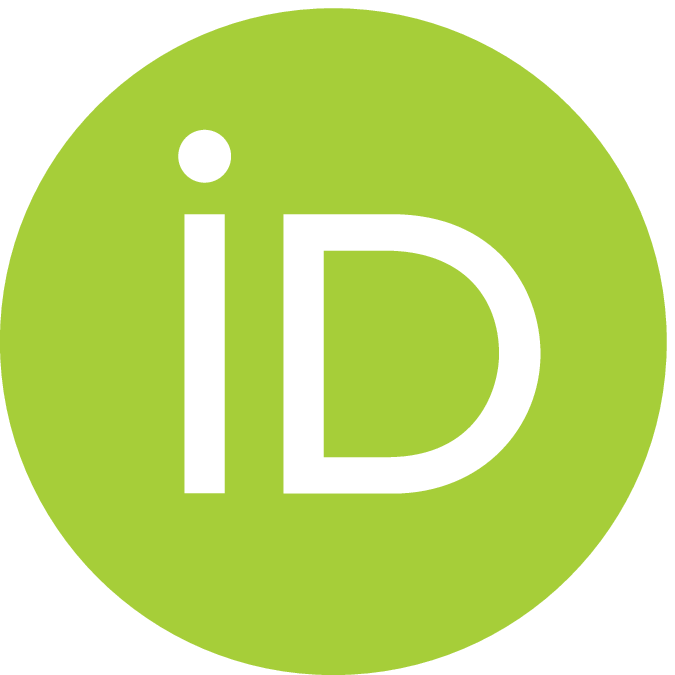}}}

\makeatletter
\renewcommand\subsubsection{\@startsection{subsubsection}{3}{\z@}%
                                     {-3.25ex\@plus -1ex \@minus -.2ex}%
                                     {-1.5ex \@plus -.2ex}
                                     {\normalfont\normalsize\bfseries}}
\makeatother

\usepackage[
    backend=biber,
    doi=true,
    sorting=none
]{biblatex}
\begin{document}
	\pagestyle{plain}
	\pagenumbering{arabic}
\title{An acrylic assembly for low temperature detectors}

\author{M. Biassoni\inst{1}
\and C. Brofferio\inst{1,2}
\and M. Faverzani\inst{1,2}
\and E. Ferri\inst{1,2}
\and S. Ghislandi\inst{2}\footnote{e-mail: \href{mailto:stefano.ghislandi@gssi.it}{stefano.ghislandi@gssi.it} (corresponding author).}\footnotemark\orcid{0000-0003-0232-1249}
\and S. Milana\inst{3}
\and I. Nutini\inst{1,2}
\and V. Pettinacci\inst{3}
\and S. Pozzi\inst{1,2}
\and S. Quitadamo\inst{2}\footnotemark[\value{footnote}]\footnotetext{Gran Sasso Science Institute, L'Aquila, Italy (present affiliation).}
}
\institute{INFN – Section of Milano Bicocca, Milan, Italy \and Department of Physics, Università degli Studi di Milano - Bicocca, Milan, Italy \and INFN - Section of Roma1, Roma, Italy}
\date{Received: date / Revised version: date}

\abstract{
Thermal detectors are a powerful instrument for the search of rare particle physics events. Inorganic crystals are classically used as thermal detectors held in supporting frames made of copper. In this work a novel approach to the operation of thermal detectors is presented, where TeO$_2$ crystals are cooled down to $\sim\SI{10}{mK}$ in a light structure built with plastic materials. The advantages of this approach are discussed.
}

\maketitle
\section{Introduction}
\label{intro}
Thermal detectors~\cite{Enns,Pirro} are largely used in rare event particle physics in the search of neutrinoless double beta decay ($0\nu\beta\beta$)~\cite{FIORINI,review} and dark matter interactions~\cite{reviewDM}. They are employed as calorimeters, thus measuring the energy released by particles, in a setup cooled down to $\sim\SI{10}{mK}$. 

One of the main experimental challenges of this technique is the capability of cooling down several crystals to extreme temperatures, maintaining them thermally stable. Detector holders characteristics are crucial to achieve satisfying performances. For this reason, copper holders have always been used for their known mechanical response at milli-Kelvin scale and high thermal conductivity. This allows to cool the material more efficiently and to provide an ideal thermal bath in which the detector discharges the absorbed energy.
The \textsf{CUORE} experiment~\cite{CUORE} succeeded in cooling and operating a \SI{1}{tonne} detector made of \ce{TeO_{2}} crystals, by building the most powerful cryostat ever made~\cite{Criostato}. 

In experiments employing \ce{^{130}Te} as $\beta\beta$ emitter, the largest gamma-background component comes from high-energy photons undergoing Compton scattering in passive elements of the system~\cite{BackgroundBudget}. Copper is therefore not an optimal choice from this point of view as it has a large Compton scattering cross-section, on top of being heavy and expensive when produced with high radio-purity standards.

We decided to test another material to be used as holder in the close vicinity of thermal detectors. 
We selected an organic compound, commercially available as Stratasys VeroClear\texttrademark.
It is a transparent PolyJet photopolymer for clear acrylic simulation. It is also inexpensive, light (density $\rho = \SI{1.18}{g/cm^{3}}$) and, being organic, characterized by a low $Z_{\text{eff}} = 6.5$\footnote{The effective atomic number has been computed as $Z_{eff} = \sqrt[2.94]{\sum_{i}^{elements} f_{i} \cdot Z_{i}^{2.94}}$ as in~\cite{MURTY1965}.}.
These features could passively lower the background induced by Compton interaction with the holder itself, meeting the requirements of rare events physics.

From a mechanical point of view, at room temperature, the VeroClear\texttrademark \,is characterized by good rigidity and strength\footnote{The VeroClear\texttrademark \,specifics are available at \url{https://www.stratasys.com/-/media/files/material-spec-sheets/mds_pj_veroclear_0320a.pdf}.}, not far from the ones in copper. Conversely the acrylic has lower elasticity. These characteristics makes the polymer suitable to build structural frames.
If copper properties are well known at cryogenic temperatures, VeroClear\texttrademark \,has to be tested in such extreme conditions.

\begin{figure}
	\begin{minipage}[t]{0.48\textwidth}
		\centering
		\includegraphics[width=0.9\linewidth]{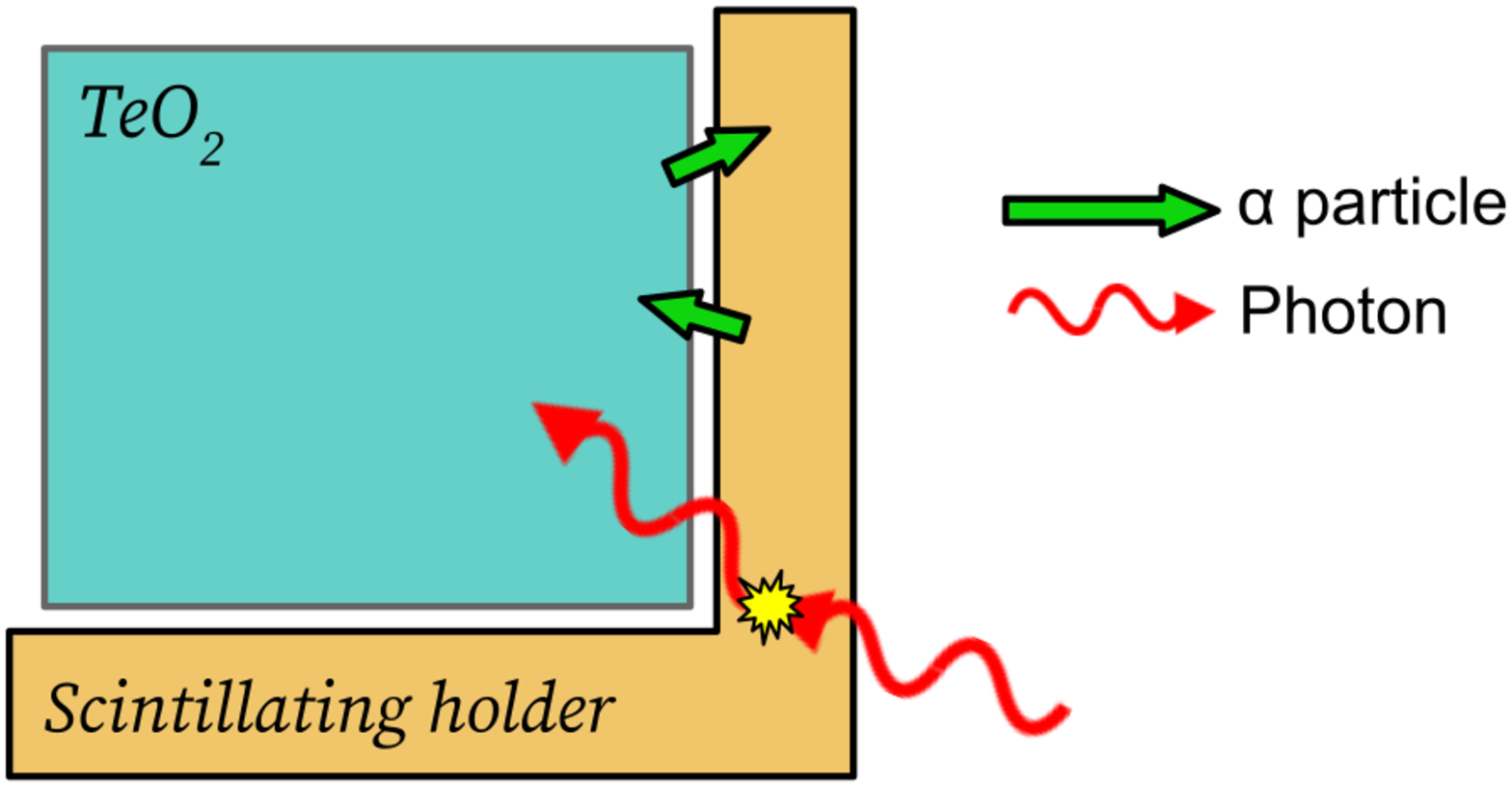}
		\caption{Sketch of the event types that contribute to the $0\nu\beta\beta$ background in \ce{TeO_{2}} and that would be vetoed by an active holder. The green straight arrows represent $\alpha$ particles generated in the holder surface, later reaching the crystals and vice-versa. The curled red arrow shows an energetic photon undergoing a Compton scattering with the holder and being absorbed by the detector}
		\label{fig:1}
	\end{minipage}\hfill
	\begin{minipage}[t]{0.48\textwidth}
		\centering
		\includegraphics[width=0.7\linewidth]{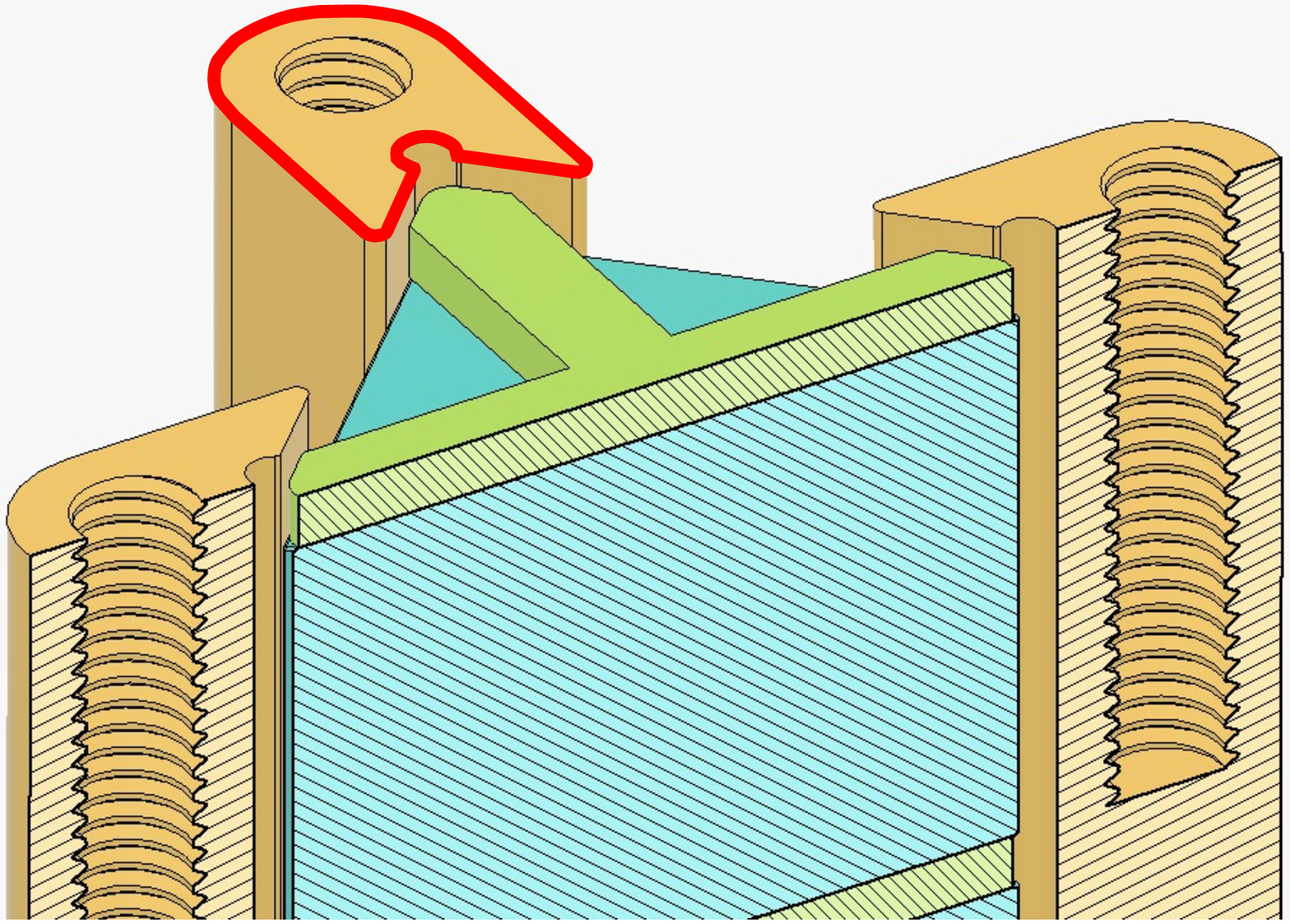}
		\caption{CAD section of the VeroClear\texttrademark \,holder together with a crystal and the relative spacer. The top face of one of the columns is highlighted in order to appreciate its shape}
		\label{fig:2}
	\end{minipage}
\end{figure}

VeroClear\texttrademark \,was born to simulate the optical properties of polymethyl methacrylate (PMMA), largely used as base for many organic scintillators~\cite{PMMAScintillator}. This opens the possibility to activate a component of the setup which is passive in classical designs. An active holder would veto its own $\alpha$ surface contamination and part of the crystals one, together with the $\gamma$ component that Compton scatters with the holder itself (See Figure~\ref{fig:1}).

The main difference of VeroClear\texttrademark \, with respect to PMMA is the possibility to be 3D printed, a useful opportunity for building structures with complex geometries, if required. Right now, we are not yet considering the impact of such a build-up technique from the point of view of induced contamination.

However, there is poor literature about properties of acrylic and, more generally, organic compounds, at the milli-Kelvin scale, even if some can still be found~\cite{Ventura}. We don't know how the VeroClear\texttrademark \,characteristics, such as low thermal conductance and large heat capacity, would modify the cool-down as well as the detector response's shape and performance.  For these reasons, an initial proof of principle is of primary importance. It is essential to check detector behaviour at cryogenic temperature, to be compared to the one coming from copper holders.


\section{Experimental Setup}
\label{sec:2}
The measurements have been performed using $1\times 1 \times 1$ \si{cm^{3}} \ce{TeO_{2}} crystals. They have been operated in a cryogenic facility located at Milano\,-\,Bicocca University (Italy). 
The crystals were equipped with $3 \times 3 \times 1$ \si{mm^{3}} Neutron Transmutation Doped \ce{Ge} (NTD) employed as thermistors~\cite{Haller}. The NTDs were mechanically and thermally coupled to the crystals with six spots of epoxy glue.
The NTD consists of a \ce{Ge} chip with a doping level slightly below the Mott transition~\cite{MottTransition}. Since the NTDs are naturally well-compensated semiconductors (the
uniformly-distributed dopants are both donors and acceptors), the \ce{Ge} lattice electrons behavior can be well described by the Shklovskii’s and Efros’ Variable Range Hopping (VRH) model~\cite{VariableRangeHopping}. In this regime, the NTD's conductivity strongly depends on the temperature $T$. It can be shown that the resistance $R$ can be written as:
\begin{equation}
    R = R_{0} e^{\sqrt{\frac{T_{0}}{T}}}.
\end{equation}
The employed NTDs had already been characterized~\cite{ArticoloSimone} and their parameters, $R_{0}$ and $T_{0}$, were respectively evaluated as $\sim$\SI{0.5}{\ohm} and $\sim$\SI{4.75}{\kelvin}. A biasing and read-out circuit provided the polarization current to the thermistors, using either \SI{20}{\mega\ohm} or \SI{15}{\giga\ohm} load resistors, depending on the NTD's resistance value at the chosen working point. The latter is defined by the position in the NTD's I-V curve, usually referred as \emph{load curve}.
The working point mainly determines the signal-to-noise ratio and the pulse shape of the detector.
\begin{figure}
    \centering
    \includegraphics[width=0.35\linewidth]{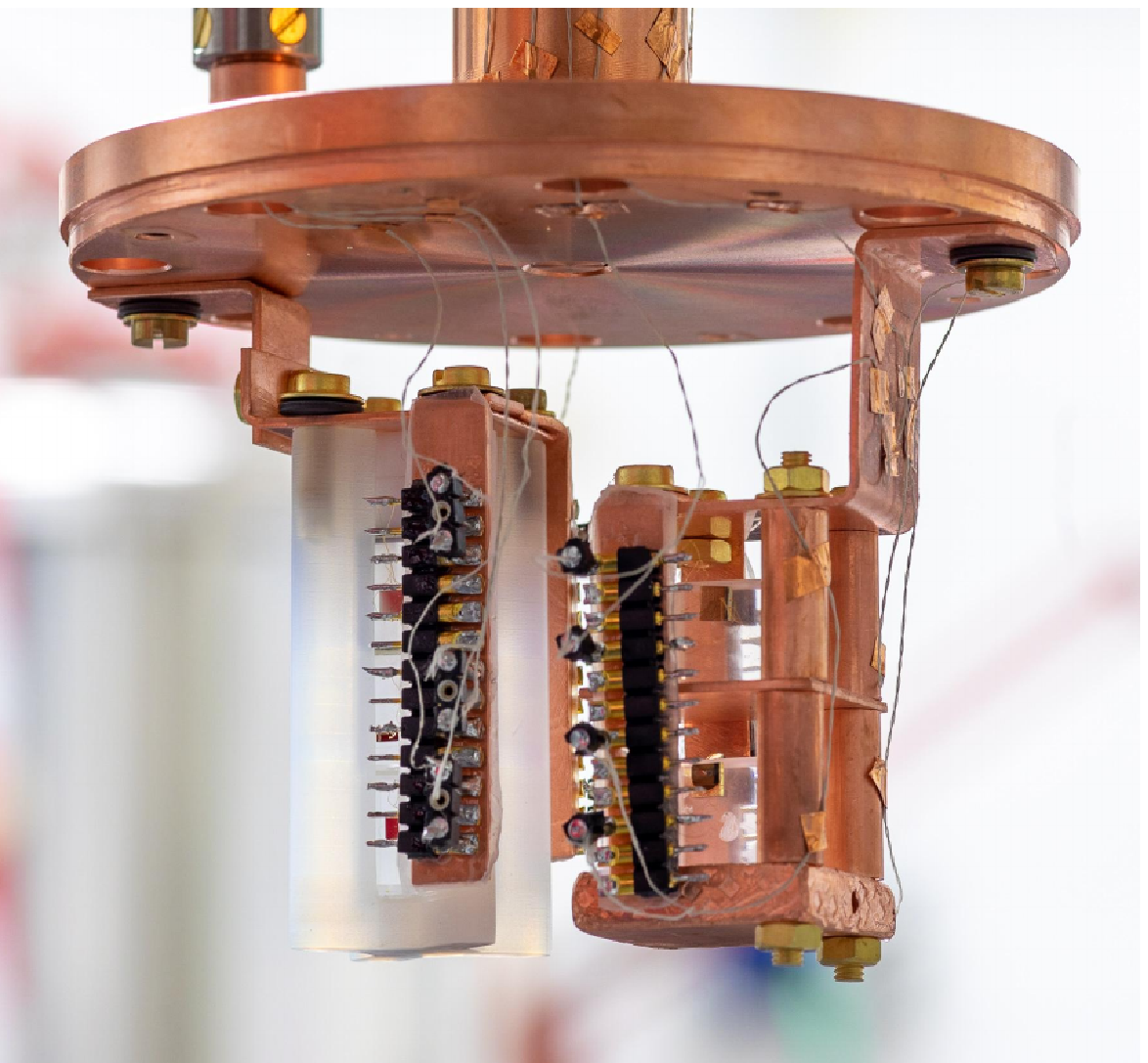}\hspace{0.2cm} \includegraphics[width=0.46\linewidth]{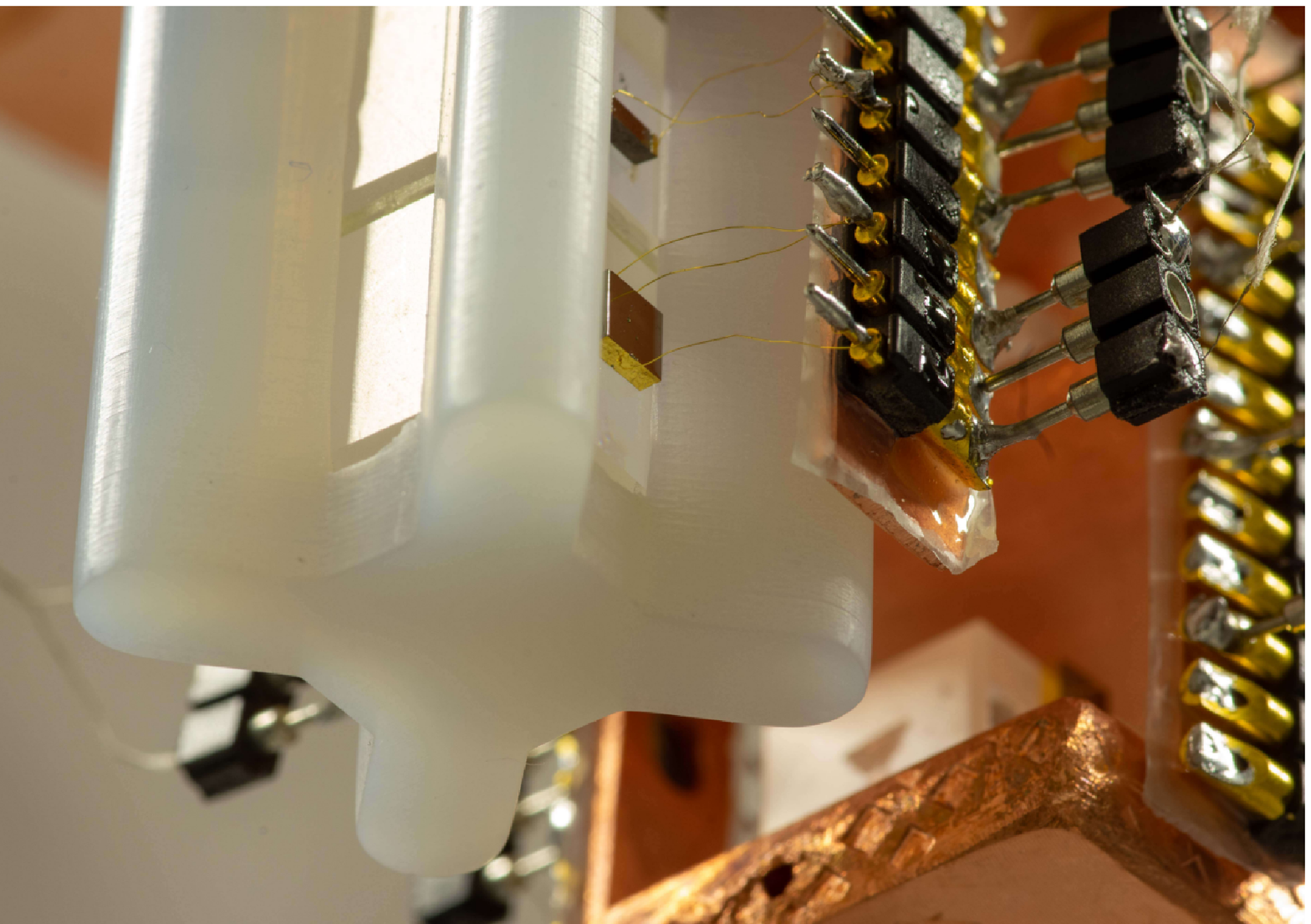}   
    \caption{Picture of the setup containing the two towers wired and installed below the cold plate through S-shaped copper strips (\emph{left}). Detail of the acrylic holder (\emph{right}). It is possible to see the NTDs glued on the crystals and their gold wire link to the Cu strips of the pins. These strips are not thermally coupled to the crystals. Only two channels for acrylic holder and one reference channel in the copper were used for this paper purposes}
    \label{fig:3}
\end{figure}

Two similar holding structures have been built for our purposes.
One is the acrylic tower, composed of a \SI{5}{mm} thick base and four columns which accommodate three \ce{TeO_{2}} stacked crystals.
To reduce the risk of damage to the detectors, due to the distinct differential contractions of acrylic and \ce{TeO_{2}}, the holder was designed to minimize stresses while maintaining the needed stiffness. This has been accomplished by shaping the VeroClear\texttrademark \,columns as in Figure~\ref{fig:2}. The holder fits the crystals' faces so to avoid vibration but leaving a mechanical unloading at the edges. Moreover, on top of the crystal stack, some space has been left to account for contractions at cryogenic temperatures.
Lastly, \SI{0.5}{mm} thick VeroClear\texttrademark \,spacers were interposed between crystals mainly to avoid thermal cross-talk among detectors.
The VeroClear\texttrademark \,holder and spacers have been produced through a Stratasys ObjetPRO machine, a 3D-Printer working by means of polyjet technology which guarantees high accuracy (order of hundredths of a mm) and quality of the parts produced through a process of instant cure by UV light of photosensitive resins (such as VeroClear\texttrademark). Soluble support material (Stratasys SUP706-B\texttrademark) was used for the construction of these parts, in order to optimize the cleaning and consequently the surface finishing. The printed substrates can be also treated for a few hours inside a lighting chamber equipped with fluorescent lamps, selected both in power and color temperature, in order to improve the final transparency. The production took place at INFN Rome mechanical workshop, exploiting the HAMMER\footnote{For more information consult \url{https://hammer.lngs.infn.it/}.} facilities (Hub for Additive Manufacturing Materials Engineering and Research) instituted by INFN. 

The second holder, made of copper, includes a \SI{5}{mm} thick base, $2$ columns and a thin plate. It can house two crystals glued, by means of Araldite$^{\circledR}$ Rapid glue spots, on the base and on the plate. 

Both the towers were hosted in an Oxford TL-200 \ce{^{3}He}-\ce{^{4}He} dilution refrigerator. 
Each holder was fixed through brass screws to a copper S-shaped strip, in turn thermally coupled to a copper plate. The latter is directly connected to the cold finger and the mixing chamber.
The cryostat operating temperature was maintained at $\sim$\SI{15}{mK} and the heat bath was monitored through a \textsf{MAGNICON} MFFT SQUID thermometer, positioned under the mixing chamber, and thermally linked to the same copper plate the detectors were connected to.
The entire detector setup is shown in fig.~\ref{fig:3}.
The NTDs readout was performed through \SI{50}{\micro m} gold wires ball-bonded to the \ce{Ge} thermistors and soldered to metallic pins on the other side. In the case of acrylic holder, these gold wires provided the detector main thermal coupling, as demonstrated in~\cite{ArticoloSimone}. The electric signal was carried out of the cryostat through twisted constantan wires, lastly reaching a Faraday cage where it underwent amplification and anti-aliasing Bessel filtering. Eventually, the detector signal was digitized and stored using two \SI{18}{bit} ADC boards.

A 2 cm thick Cu layer surrounded the setup to reduce the environmental $\gamma$ radiation reaching the two towers, together with a \SI{5}{cm} thick Pb disk on top of the cold plate.
A \ce{^{232}Th} source was employed to calibrate the energy scale and was put under the cryostat during calibration runs. 

\section{Data analysis}
\label{sec:3}
When a particle passes through the \ce{TeO_{2}} crystals, it releases energy and, consequently, the detector temperature raises. This variation is detected through the NTD, operated as a thermometer.

\begin{figure}
	\begin{minipage}[t]{0.48\textwidth}
		\centering
		\includegraphics[width=1\linewidth]{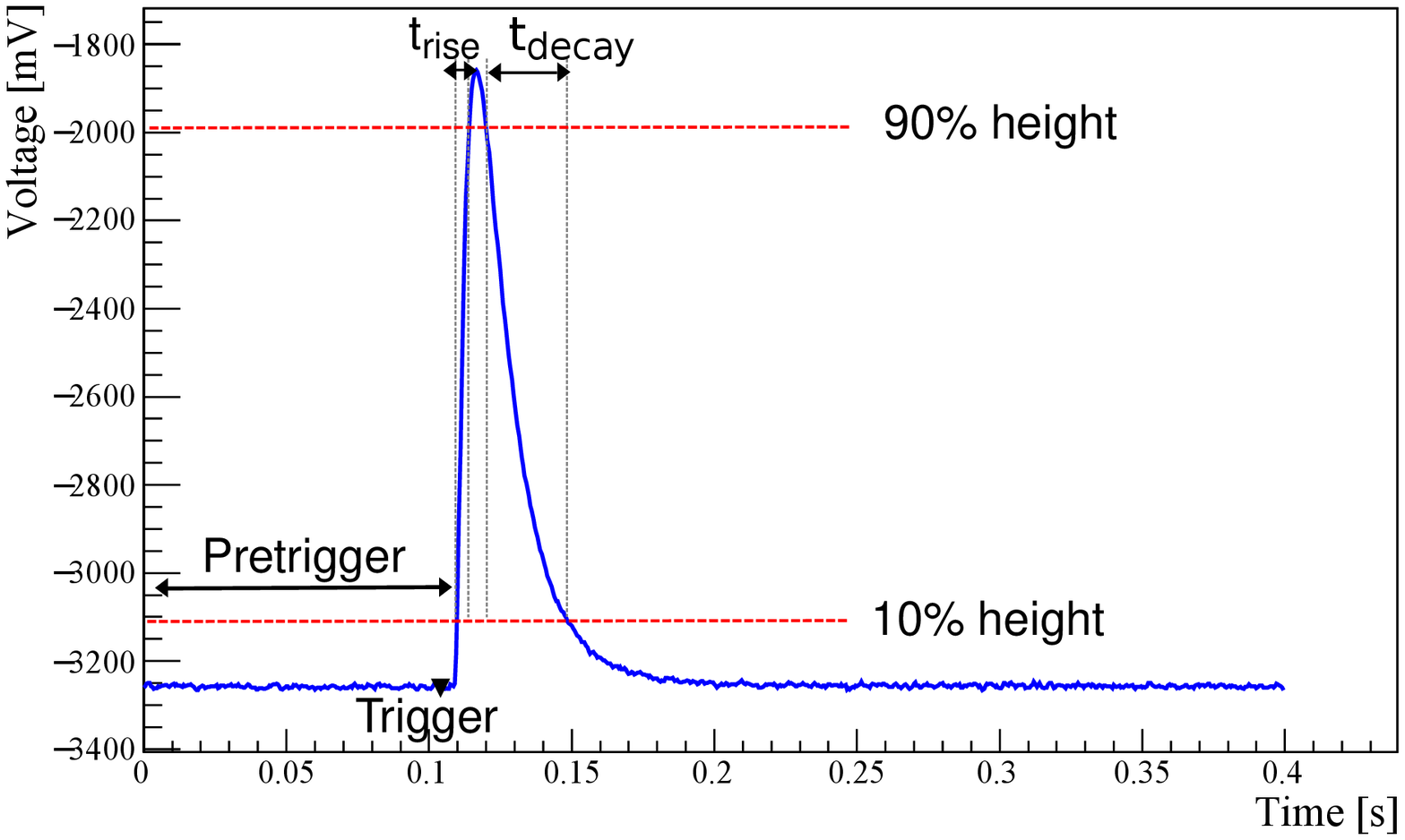}
		\caption{Typical \SI{400}{\milli\second} event window containing a pulse. Rise time ($t_{\text{rise}}$) and decay time ($t_{\text{decay}}$) are defined as the time elapsed between $10\%$ and $90\%$ of the pulse amplitude}
		\label{fig:4}
	\end{minipage}\hfill
	\begin{minipage}[t]{0.48\textwidth}
		\centering
		\includegraphics[width=1.0\linewidth]{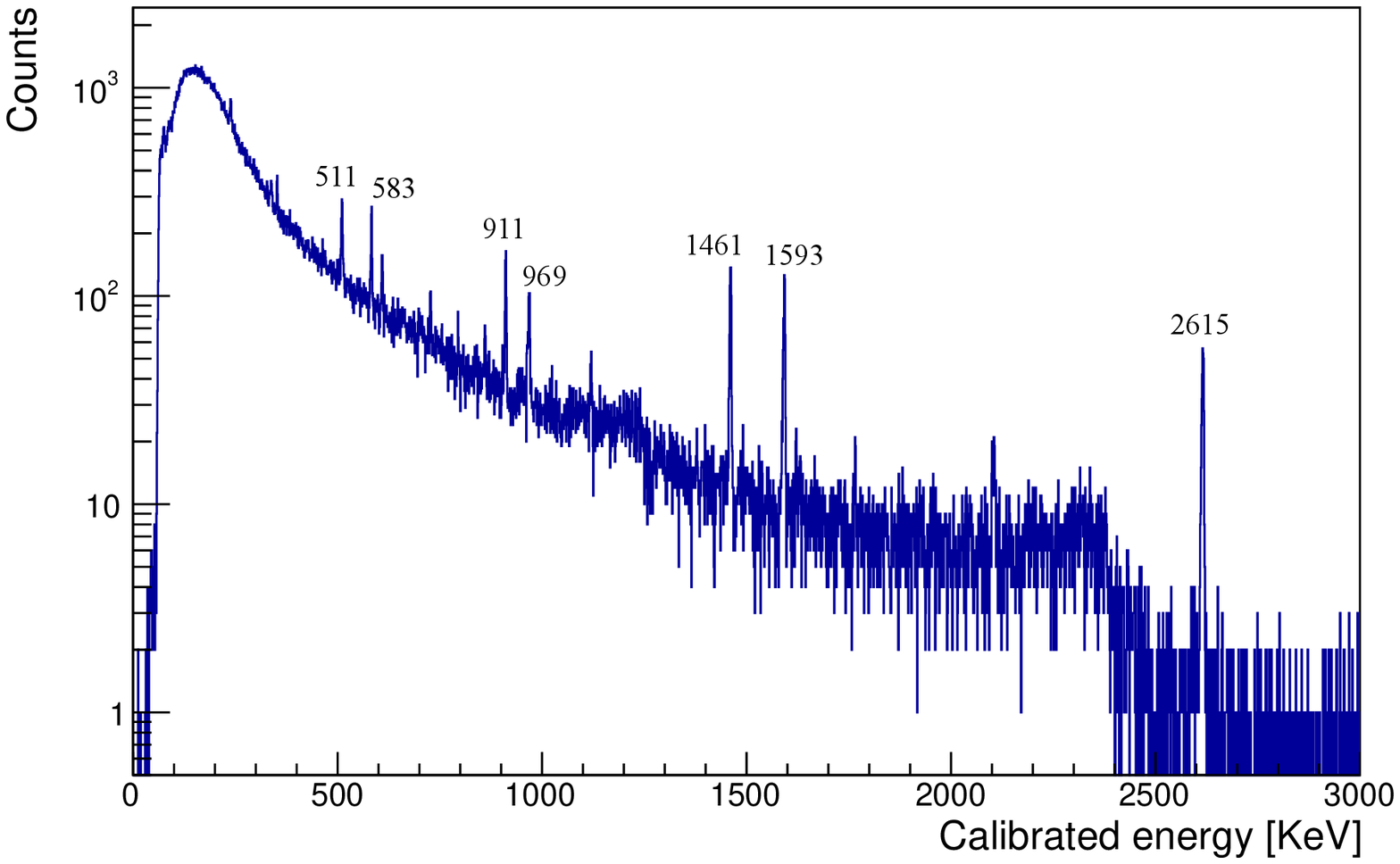}
		\caption{Example of calibration energy spectrum. The energy of lines used for the calibration is reported on top of the relative peaks and expressed in \si{keV}. The \SI{511}{keV} peak is due to $e^{+}$-$e^{-}$ annihilation, \SI{583}{keV} and \SI{2615}{keV} belong to \ce{^{208}Tl}, \SI{911}{keV} and \SI{969}{keV} come from \ce{^{228}Ac} decay, the \SI{1461}{keV} peak is due to \ce{^{40}K} and the \SI{1593}{keV} is the double escape peak of the highest energy \ce{^{208}Tl} line}
		\label{fig:5}
	\end{minipage}
\end{figure}

The NTDs output was digitized and stored as a continuous \SI{2}{kHz} waveform.
During the data-taking, a derivative trigger flagged events from the continuous stream. The selected pulses were stored as windows of \SI{400}{ms} length (Figure~\ref{fig:4}). The pulse window contains a \SI{100}{ms} pre-trigger region used to extract information about the specific event baseline.
Moreover, noise windows were sampled randomly. They have been used later to construct a noise power spectrum for each channel, run by run.

The first part of the offline analysis identified periods of time in which the detectors were misbehaving. The check was done by looking for intervals with unstable baseline or high baseline RMS. These ``bad intervals'' were excluded from the subsequent analysis steps.
For each channel the \emph{average pulse} (AP) and the \emph{average noise power spectrum} (ANPS) have been computed. They were used to evaluate the pulse amplitude with the Optimum Filter technique~\cite{OptimumFilter}, which allows to maximize the signal-to-noise ratio.

Since the thermal gain depends on temperature, it was essential to correct the detector response compensating for temperature variations among events. For this reason, we applied a thermal gain stabilization by using the detector baseline level, that is the pretrigger average baseline, as a temperature proxy.
The temperature dependence of the gain was corrected by correlating the amplitude of pulses belonging to a peak of known energy (\SI{2615}{keV} from \ce{^{208}Tl}) to the baseline level~\cite{ThermalStabilization}.

The spectrum was calibrated using recognizable $\gamma$ peaks, mainly coming from the \ce{^{232}Th} calibration source. These lines have been fitted using a Gaussian function added to a linear background. For the calibrating function a $2^{nd}$ order polynomial with intercept at zero has been employed. An example of the obtained energy spectra is reported in fig.~\ref{fig:5}.

All considered, we analyzed one reference channel in the copper assembly (here named channel 1) and $2$ channels from the acrylic tower (channel 2 and channel 3).

\section{Results}
\label{sec:4}
During the cool-down some high-statistics calibration runs were collected. They all lasted more than \SI{20}{hr} and they contain a total of $~10^{5}-10^{6}$ physical pulses each. Before the data acquisition, test runs were needed to optimize NTDs' working point and trigger parameters. 

\begin{figure}
    \centering
    \includegraphics[width=0.6\linewidth]{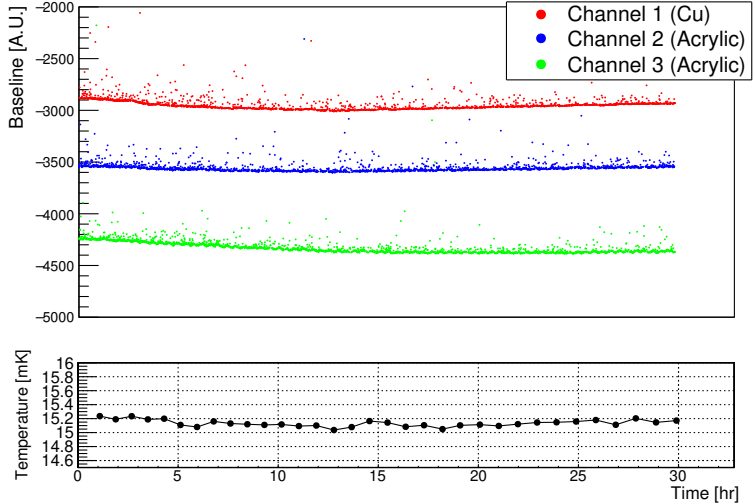}
    \caption{Baseline level-time scatter plot for channels $1$, $2$ and $3$ of a sample run. Arbitrary offsets have been added to not overlap all the points. The baseline, proxy of the temperature at the crystal level, is stable over the \SI{30}{hr} of data-taking both for acrylic and copper towers. The detectors in the VeroClear\texttrademark \, don't show long time shifts or spikes out of the expected ones. The visible overall baseline modulation, present in all the three channels, is due to fluctuations of the cryostat temperature, reported in the lower plot} 
    \label{fig:6}
\end{figure}

The first obtained result is related to the detector stability over time. 
Acrylic is amorphous and it has low thermal conductance. The disordered molecule structure could in principle relax and, unpredictably, release energy, acting as a residual heat load on the holder. Therefore, the presence of acrylic could induce temperature shifts on long time scales or fast and large spikes. The latter, in contrast to the usual temperature variations, could not be corrected with the stabilization described in the previous chapter.
However, the acrylic setup baseline level, proxy for the crystals temperature in the holder, was as stable as the one in the copper holder for very long times. The stability has been checked for all channels and all acquired runs. An example is shown in fig.~\ref{fig:6}.

Since crystal holders have also structural roles, another important result concerns acrylic mechanical properties.
The plastic holder had already undergone many cool-down cycles from room temperature to $\SI{77}{K}$ and few cycles from room temperature to $\sim \SI{15}{mK}$. After each group of tests it was examined carefully at the microscope and no cracks or visible defects were found. No spurious thermal noise of the type registered in CRESST~\cite{CRESST} was observed during data taking, nor cracks on the crystals after warm-up.

\begin{figure}
	\begin{minipage}[t]{0.48\textwidth}
        \centering
        \includegraphics[width=0.97\linewidth]{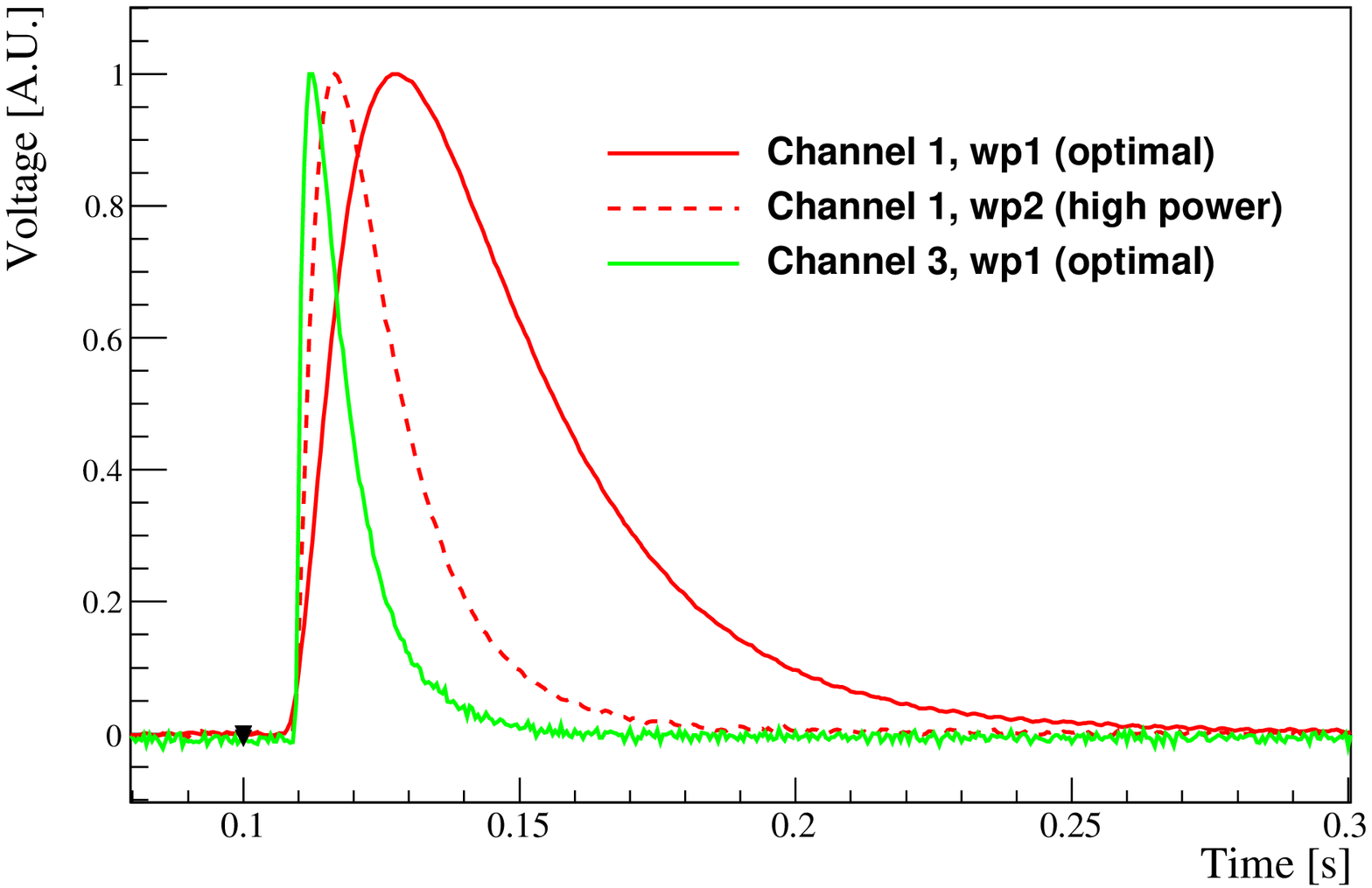}
		\caption{Examples of pulses coming from detectors in acrylic holder at \emph{wp1}, the optimal working point and from copper assembly operated at \emph{wp1} and \emph{wp2}, obtained by dissipating \SI{30}{pW} on channel 1. Notice that this is not the entire \SI{400}{ms} acquired time window}
        \label{fig:7}
	\end{minipage}\hfill
	\begin{minipage}[t]{0.48\textwidth}
		\centering
		\includegraphics[width=0.97\linewidth]{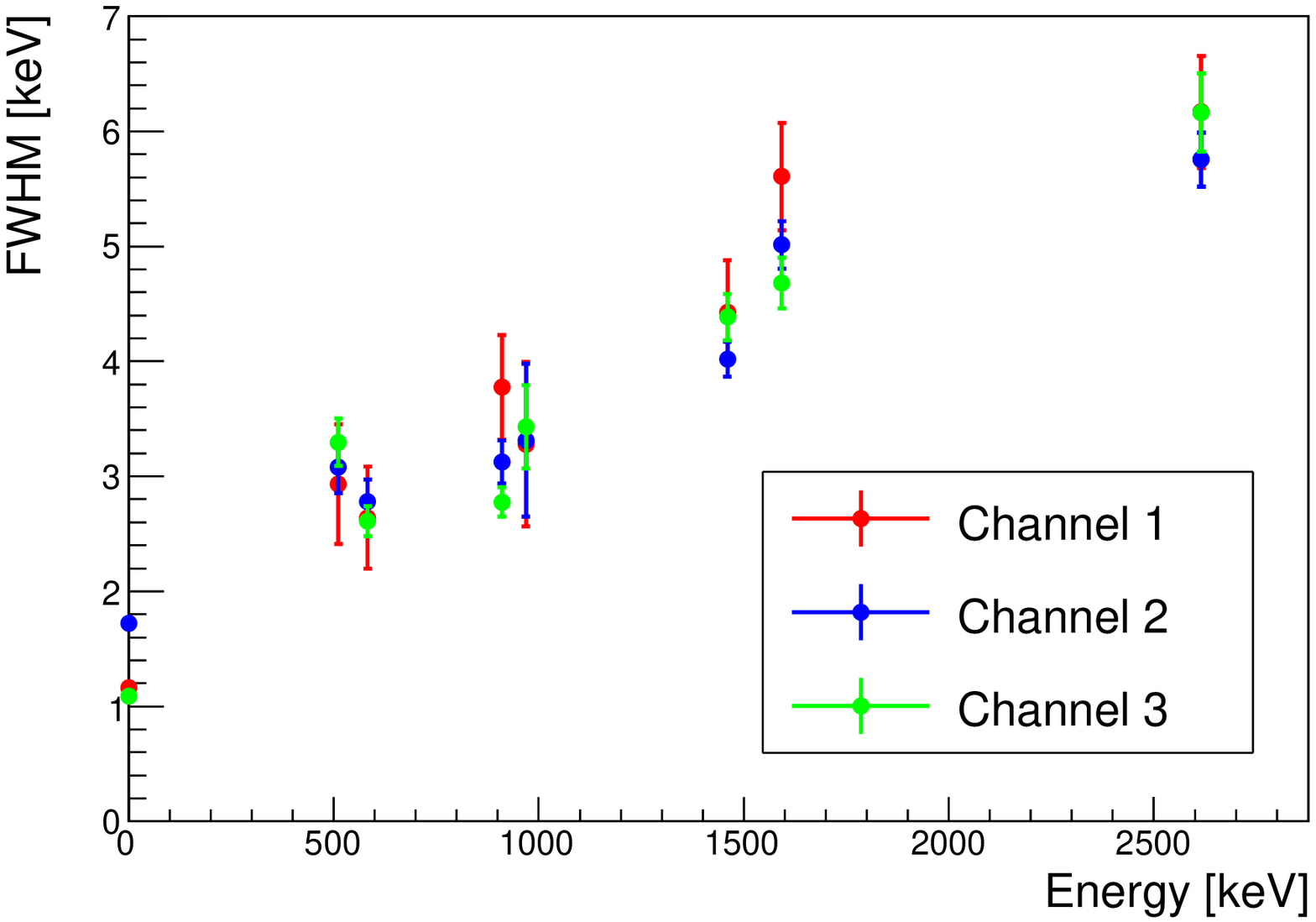}
        \caption{Energy resolution expressed as FWHM for the considered channels. The point at zero energy is the resolution of randomly acquired noise windows}
		\label{fig:8}
	\end{minipage}
\end{figure}

We expected the different conductivity of the holding material~\cite{ArticoloSimone} to influence pulse formation. Indeed, at their respective optimal working points (\emph{wp1}) the signals from channel in the acrylic holder are faster than those from channel $1$ (Cu tower). The effect can be qualitatively appreciated by comparing the pulse shapes plotted in fig.~\ref{fig:7}. This can be explained by the weaker thermal coupling provided by the acrylic, leading to an higher crystals operating temperature of $\sim \SI{20}{mK}$ (with smaller working resistances $R_W$ resulting in smaller intrinsic gain but faster time response). Conversely, the detectors in the copper tower were at the cryostat temperature, that was $\sim \SI{15}{mK}$. Since fast pulses could be useful to reduce pile-up whenever this background is an issue (as in the proposed CUPID experiment~\cite{CUPIDpileup, CUPIDpileup2}), this acrylic holder additive feature is very promising. To achieve similarly fast pulses on standard Cu frames, the NTDs can be warmed up by increasing the bias current. This approach was applied to channel $1$ by operating it at a different working point (\emph{wp2}), dissipating $\sim$\SI{30}{pW} through the bias circuit\footnote{In the case of \emph{wp1} the dissipated power is negligible, in the order of \SI{0.5}{pW}.}. The dissipated power was such that the pulses duration were shortened but the resolution (and the pulse shape) didn't suffered the non optimal working point yet.
We characterize the signals by means of their rise and decay time, defined as the time elapsed for a pulse from $10\%$ to $90\%$ of its amplitude and vice versa. The obtained average values are reported in table~\ref{tab:1}, together with their working resistances $R_{\text{w}}$. Using these results, the pile-up probability, modelled as a Poissonian distribution, would be lower in acrylic holder channels by a factor $\sim 5$, when the NTDs are operated at their optimal working point.
\begin{table}
    \centering
    \caption{Average rise time and decay time, as defined in fig.~\ref{fig:4}, computed for the two mentioned working points: \emph{wp1} and \emph{wp2}. In the last column, the NTDs working resistances.}
    \begin{tabular}{cccccc}
    \hline\noalign{\smallskip}
    Holder material & Channel & Working point & Rise time [\si{\milli\second}] & Decay time [\si{\milli\second}] & NTD $R_{\text{w}}$\\
    \noalign{\smallskip}\hline\noalign{\smallskip}
    Copper & 1 & \emph{wp1} & $24.0\pm 0.2$ & $64.8\pm 0.8$ & $\SI{12.8}{\mega\ohm}$\\
    Acrylic & 2 & \emph{wp1} & $4.4 \pm 0.1$ & $47.6 \pm 0.5$ & $\SI{139}{\kilo\ohm}$\\ 
    Acrylic & 3 & \emph{wp1} & $2.49 \pm 0.05$ & $17.8 \pm 0.1$ & $\SI{287}{\kilo\ohm}$\\
    Copper & 1 & \emph{wp2} & $4.6 \pm 0.1$ & $30.4 \pm 0.4$ & $\SI{1.37}{\mega\ohm}$\\
    \noalign{\smallskip}\hline
    \end{tabular}
	\label{tab:1}
\end{table}
Besides these thermal considerations, the pulse shape of the detectors housed in the acrylic holder is not further modified. This fact allowed to process events from the two setups with the same algorithms without adapting or tuning the employed software.

An important parameter used to assess the detector performance is the FWHM energy resolution. The most prominent calibration spectrum peaks, whose energies are reported in fig.~\ref{fig:5}, have been fitted with a Gaussian function plus a linear background at the end of the calibration step. A zero-energy FWHM, corresponding to the baseline resolution, has been computed taking into account the baseline distribution spread evaluated after filtering.
The obtained resolution-energy plot is shown in fig.~\ref{fig:8}. The resolution is dominated, at low energies, by the baseline noise. Then, it grows monotonically, following the typical low temperature calorimeters trend.
The FWHM resolutions of baseline and of the 
highest energy peak of the calibration are reported in table~\ref{tab:2}, where channel 1 is operated at \emph{wp2} and channel 2 and 3 at \emph{wp1}.

\begin{table}
    \centering
    \caption{Summary of the results, in terms of energy resolution, obtained for the considered channels. In the table, the baseline FWHM resolution and the \SI{2615}{keV} peak one are reported.}
    \begin{tabular}{rccc}
    \hline\noalign{\smallskip}
    Channel & Holder material & FWHM baseline [\si{keV}] & FWHM $@$ \SI{2615}{keV} [\si{keV}] \\
    \noalign{\smallskip}\hline\noalign{\smallskip}
    $1$ & Copper & $1.31 \pm 0.02$ & $6.1\pm0.5$ \\
    $2$ & Acrylic & $1.72 \pm 0.02$ & $5.3\pm 0.3$ \\
    $3$ & Acrylic & $1.04 \pm 0.01$ & $6.1\pm 0.4$ \\
    \noalign{\smallskip}\hline
    \end{tabular}
	\label{tab:2}
\end{table}

\begin{figure}
    \centering
    \includegraphics[width=0.6\linewidth]{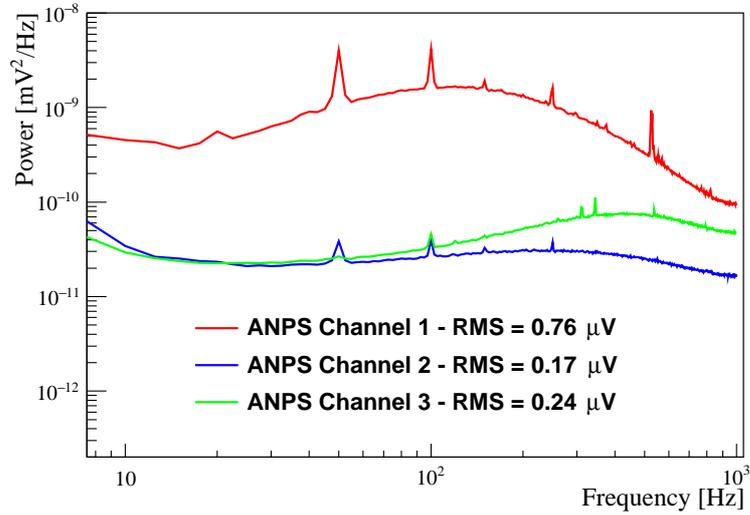}
    \caption{Normalized ANPS for channel $1$, belonging to the copper tower, operated at \emph{wp2} and channels $2$ and $3$, framed in the acrylic holder and operated at \emph{wp1}. The difference in the amplitude scale and in the RMS can be ascribed to the different $R_W$ of the detectors (see tab~\ref{tab:1}), contributing to the Johnson noise on the channels}
    \label{fig:9}
\end{figure}

We also compared the noise registered on the different channels by computing, run by run, their ANPS and normalizing for the electronic gain. As can be seen from fig.~\ref{fig:9}, the noise belonging to channels in the acrylic tower doesn't show problematic features when operated at the optimal working points.
As already pointed out, the crystals in the acrylic work at slightly higher temperatures, that is at lower $R_{w}$. As a check, the Johnson contribution to the continuum is confirmed to be lower.

\section{Conclusions and Comments}
\label{sec:5}
Neutrinoless double beta decay searches, decade after decade, improve the limits on the $0\nu\beta\beta$ half-life for different emitting isotopes.
Future experiments will ask for more demanding material radio-purity, background rejection capability, and better energy resolutions. For this reason, acting on structures close to the detectors will be essential to fulfill the desired requirements.
In the field of low temperature detectors, copper has always been employed for the construction of the crystal holding structures. However, copper is expensive and its Compton scattering cross-section can cause significant background for experiments using \ce{^{130}Te} as target isotope.
A novel approach would consist in using holding structures made of light organic compounds, which would reduce both the cost and the impact on the measured background.

In our experimental setup, VeroClear\texttrademark \,has shown good mechanical characteristics and thermal coupling with our crystals down to the milli-Kelvin scale. Since it can be 3D printed, even complicated structures can be realised with a low number of components, simplifying the assembly. 
From a mechanical point of view, it would be possible to scale our test holder to real experiments dimension. 
A VeroClear\texttrademark \,holder would bear the \textsf{CUORE} detector strings, while maintaining the same frame volume employed in the experiment (with great margin, thus leaving room for further optimization).
There are already examples of acrylic laminates produced with low \ce{^{238}U} and \ce{^{232}Th} contamination, at levels comparable to the \textsf{CUORE} cleanest copper.
Therefore, the bulk contamination introduced by the detectors' holder would gain a factor at least 7.5 (density ratio), passively lowering the induced background. 

We demonstrated that the material replacement doesn't compromise the detector working properties (resolution, stability, etc.). The detectors housed in the plastic tower are as stable as the ones framed in copper for long times. 
In addition, their equilibrium temperature is only a few \si{mK} higher than in the copper case. As a consequence, the pulses are faster, allowing to reduce the background induced by $2\nu\beta\beta$ pile-up in $0\nu\beta\beta$ searches with isotopes characterized by shorter half-life~\cite{CUPIDpileup,CUPIDpileup2}. The acrylic holder doesn't cause any deterioration in the energy resolution and in the noise power spectrum.
Since the acrylic frame is characterized by low conductance, thermalization issues cannot be excluded, especially when the detector strings become longer. These kind of tests are required to check where such effects are negligible and where they become a severe issue.

A first hint of a real reduction of the Compton background events has been obtained and validated by Monte Carlo simulations and it is discussed in a specific paper~\cite{ArticoloSIF}. We will therefore further proceed in this direction in the near future, studying the VeroClear\texttrademark \,scintillation light properties at the milli-Kelvin scale, together with the associated readout. These tests will be critical to evaluate the feasibility of an ``active" acrylic holder, useful to reach a further background rejection mainly tagging and suppressing $\alpha$ induced events and rejecting $\beta$ radiation coming from the components close to the detectors. 
These additional features will be essential for future-generation $0\nu\beta\beta$ experiments, expected to operate close to the background-free regime.

\section*{Acknowledgments} 
We thank Marco Iannone from INFN - Section of Roma1 and member of the Hammer project for helping with the production and preparation of the 3D printed components of the assembly.

\printbibliography

\end{document}